\documentclass[sigconf,authorversion,nonacm]{acmart}

\def\BibTeX{{\rm B\kern-.05em{\sc i\kern-.025em b}\kern-.08emT\kern-.1667em\lower.7ex\hbox{E}\kern-.125emX}}

\usepackage{nicefrac}
\usepackage{siunitx}
\usepackage{array,framed}
\usepackage{booktabs}
\usepackage{
  color,
  soul,
  float,
  epsfig,
  wrapfig,
  graphicx
}
\usepackage{setspace}
\usepackage{latexsym,fancyhdr,url}
\usepackage{enumerate}
\usepackage{algpseudocode}
\usepackage{graphics}
\usepackage{xparse} 
\usepackage{xspace}
\usepackage{multirow}
\usepackage{csvsimple}
\usepackage{balance}
\usepackage{braket}
\usepackage{csquotes}
\usepackage{tabularx}
\usepackage{enumitem}
\setlist{topsep=2pt, leftmargin=*}
\usepackage{graphicx, caption, subcaption}

\usepackage{
  tikz,
  pgfplots,
  pgfplotstable
}
\usepackage{hyperref}

\usetikzlibrary{
  shapes.geometric,
  arrows,
  external,
  pgfplots.groupplots,
  matrix
}

\pgfplotsset{compat=1.9}

\usepackage{mathtools}

\usepackage{xcolor}

\DeclareMathAlphabet{\mathcal}{OMS}{cmsy}{m}{n}

\DeclareGraphicsExtensions{%
    .png,.PNG,%
    .pdf,.PDF,%
    .jpg,.mps,.jpeg,.jbig2,.jb2,.JPG,.JPEG,.JBIG2,.JB2}

\usepackage{xparse}
\newcommand{\bnm}{\begin{newmath}}
\newcommand{\enm}{\end{newmath}}

\newcommand{\bea}{\begin{eqnarray*}}%
\newcommand{\eea}{\end{eqnarray*}}%

\newcommand{\bne}{\begin{newequation}}
\newcommand{\ene}{\end{newequation}}

\newcommand{\bal}{\begin{newalign}}
\newcommand{\eal}{\end{newalign}}

\newenvironment{newalign}{\begin{align}%
\setlength{\abovedisplayskip}{4pt}%
\setlength{\belowdisplayskip}{4pt}%
\setlength{\abovedisplayshortskip}{6pt}%
\setlength{\belowdisplayshortskip}{6pt} }{\end{align}}

\newenvironment{newmath}{\begin{displaymath}%
\setlength{\abovedisplayskip}{4pt}%
\setlength{\belowdisplayskip}{4pt}%
\setlength{\abovedisplayshortskip}{6pt}%
\setlength{\belowdisplayshortskip}{6pt} }{\end{displaymath}}

\newenvironment{newequation}{\begin{equation}%
\setlength{\abovedisplayskip}{4pt}%
\setlength{\belowdisplayskip}{4pt}%
\setlength{\abovedisplayshortskip}{6pt}%
\setlength{\belowdisplayshortskip}{6pt} }{\end{equation}}

\newcounter{ctr}

\newcounter{mytable}
\def\mytable{\begin{centering}\refstepcounter{mytable}}
\def\endmytable{\end{centering}}

\newcounter{myfig}
\def\myfig{\begin{centering}\refstepcounter{myfig}}
\def\endmyfig{\end{centering}}

\newlength{\saveparindent}
\newlength{\saveparskip}
\newcommand{\E}{{\rm I\kern-.3em E}}

\renewcommand{\eqref}[1]{\mbox{Equation~(\ref{#1})}}

\def \part {part}

\renewcommand{\paragraph}[1]{\vspace*{6pt}\noindent\textbf{#1}\;}

\def \blackslug{\hbox{\hskip 1pt \vrule width 4pt height 8pt
    depth 1.5pt \hskip 1pt}}
\def \qed{\quad\blackslug\lower 8.5pt\null\par}

\newcounter{mynote}[section]

\newcommand\ignore[1]{}

\newcounter{rcnote}[section]

\newcounter{mrnote}[section]

\newcounter{fknote}[section]

\newcounter{anote}[section]

\DeclareMathSymbol{\mlq}{\mathord}{operators}{``}
\DeclareMathSymbol{\mrq}{\mathord}{operators}{`'}

\newcommand{\rhf}[2]{R_{f, \gamma}}

\DeclareDocumentCommand{\edist}{o o}{
  \ensuremath{
    \IfNoValueTF{#1}{{d}}{{\sf d}(#1,#2)}
  }
}

\newcommand{\olrk}[1]{\ifx\nursymbol#1\else\!\!\mskip4.5mu plus 0.5mu\left(\mskip0.5mu plus0.5mu #1\mskip1.5mu plus0.5mu \right)\fi}

\NewDocumentCommand{\indseq}{ O{1} O{r} }{{#1}\ldots {#2}}

\copyrightyear{2023}
\acmYear{2023}
\setcopyright{acmlicensed}\acmConference[NoCArc '23]{16th edition of International Workshop on Network on Chip Architectures}{October 28, 2023}{Toronto, ON, Canada}
\acmBooktitle{16th edition of International Workshop on Network on Chip Architectures (NoCArc '23), October 28, 2023, Toronto, ON, Canada}
\acmPrice{15.00}
\acmDOI{10.1145/3610396.3623267}
\acmISBN{979-8-4007-0307-2/23/10}

\begin{document}
\title{Interconnect Fabrics for Multi-Core Quantum Processors:\\A Context Analysis}

\author{Pau Escofet*}
\affiliation{
\institution{NaNoNetworking Center in Catalunya (N3Cat)\\Universitat Polit\`{e}cnica de Catalunya}
  \city{Barcelona}
  \country{Spain}
  }
\email{}

\author{Sahar Ben Rached*}
\affiliation{
\institution{NaNoNetworking Center in Catalunya (N3Cat)\\Universitat Polit\`{e}cnica de Catalunya}
  \city{Barcelona}
  \country{Spain}
  }
\email{}

\author{Santiago Rodrigo}
\affiliation{
\institution{NaNoNetworking Center in Catalunya (N3Cat)\\Universitat Polit\`{e}cnica de Catalunya}
  \city{Barcelona}
  \country{Spain}
  }
\email{}

\author{Carmen G. Almudever}
\affiliation{
 \institution{Computer Engineering Department\\Universitat Polit\`{e}cnica de Val\`{e}ncia}
 \city{Valencia}
 \country{Spain}}
\email{}

\author{Eduard Alarc\'on}
\affiliation{
 \institution{NaNoNetworking Center in Catalunya (N3Cat)\\Universitat Polit\`{e}cnica de Catalunya}
  \city{Barcelona}
  \country{Spain}
  }
 
\author{Sergi Abadal}
\affiliation{
 \institution{NaNoNetworking Center in Catalunya (N3Cat)\\Universitat Polit\`{e}cnica de Catalunya}
  \city{Barcelona}
  \country{Spain}
  }

\date{}

\renewcommand{\shortauthors}{P. Escofet, S. Ben Rached, S. Rodrigo, C. G. Almudever, E. Alarc\'on, S. Abadal}%
\renewcommand{\shorttitle}{Interconnect Fabrics for Multi-Core Quantum Processors: A Context Analysis}%

\begin{abstract}
Quantum computing has revolutionized the field of computer science with its extraordinary ability to handle classically intractable problems. To realize its potential, however, quantum computers need to scale to millions of qubits, a feat that will require addressing fascinating yet extremely challenging interconnection problems. In this paper, we provide a context analysis of the nascent quantum computing field from the perspective of communications, with the aim of encouraging the on-chip networks community to contribute and pave the way for truly scalable quantum computers in the decades to come. \let\thefootnote\relax\footnotetext{* Equally contributing authors. Authors acknowledge support from the European Research Council (ERC) under GA 101042080 (WINC) and the European Innovation Council (EIC) Pathfinder scheme, GA 101099697 (QUADRATURE).
This is the author's version of the work. It is posted here for your personal use. Not for redistribution. The definitive Version of Record can be found at https://doi.org/10.1145/3610396.3623267
}
\end{abstract}



\maketitle

\section{Introduction}
\label{sec:intro}
Quantum computing proposes a new paradigm for solving computational problems by leveraging fundamental properties of quantum systems, such as superposition and entanglement \cite{nielsen00}.
These properties allow quantum computers to achieve a computational speedup over classical systems in solving certain problems, some of which would be intractable otherwise. As a result, this technology offers a vast array of potential applications spanning across various fields, including prime factorization \cite{10.1137/S0097539795293172}, database search \cite{10.1145/237814.237866}, physics \cite{Low_2019}, chemistry \cite{Peruzzo_2014}, finance \cite{Woerner_2019}, and healthcare \cite{fi15030094}.


Current quantum computers use diverse qubit technologies such as superconducting qubits \cite{nakamura_1999_coherent}, photonic qubits \cite{pieter_2007_linear}, quantum dots \cite{PhysRevLettQDS} and trapped ions \cite{PhysRevLettTI}, not exceeding a few hundred of moderately robust qubits \cite{chow_2021_ibm}. Although existing roadmaps point toward processors hosting a few thousand qubits in the near future \cite{gambetta_2022_expanding} and improving the robustness of the physical qubits \cite{NAP25196, 7927104}, there is still a considerable gap towards the millions of qubits that will be needed for addressing practical real-world problems \cite{Preskill2018quantumcomputingin}.

Densely-packed monolithic quantum processors hosting a large number of qubits impose severe technical issues due to the effect of cross-talk, quantum state disturbance, and the increased complexity of the systems used to control the qubits \cite{9268630}, which deteriorates the computational results. Moreover, the interconnect between the host computer and the quantum processor (typically of extremely different form factors and placed in hugely different temperature levels) quickly becomes a bottleneck in such architectures \cite{malinowski2023wire, potovcnik2021millikelvin}. Therefore, scaling up current quantum computers to host a higher number of qubits in such monolithic architectures remains a huge challenge, and finding methods to alleviate these constraints is instrumental to developing large-scale, viable quantum computers. 

A proposed alternative to the monolithic quantum computer architecture is modular (or multi-core) quantum processors \cite{rodrigo2021double, smith2022scaling, jnane2022multicore}. This approach is based on a \emph{scale out} approach, i.e., interconnecting several moderately sized quantum processing units (QPUs) or \emph{quantum cores} via classical and quantum-coherent links \cite{gold2021entanglement}, for the purpose of mitigating the challenges associated with scaling up the number of qubits on a single chip. In this case, however, the interconnect fabric within such quantum architectures emerges as a critical sub-system as the number of quantum cores increases. 

Since the interconnects appear to be one of the key elements enabling the scaling of quantum computers, the present paper aims to provide a context analysis of the quantum computing field, intended to inspire the Network-on-Chip (NoC) community to engage with its unique communications challenges. Towards that goal, our contributions include: (i) a brief tutorial on multi-core quantum computers, describing a simplified stack from software to hardware, given in Section~\ref{sec:background}; (ii) an outline of the main communication flows in quantum computers and the different interconnect technologies that could realize them, as done in Section~\ref{sec:comms}; and (iii) an analysis of the communications context in modular quantum computers, including a characterization of the qubit traffic in such systems, as performed in Section~\ref{sec:context}. Finally, (iv) we discuss several outstanding challenges in Section~\ref{sec:challenges} and conclude the paper in Section~\ref{sec:conclusion}. 


\begin{figure}
     \centering
     \includegraphics[width=\columnwidth]{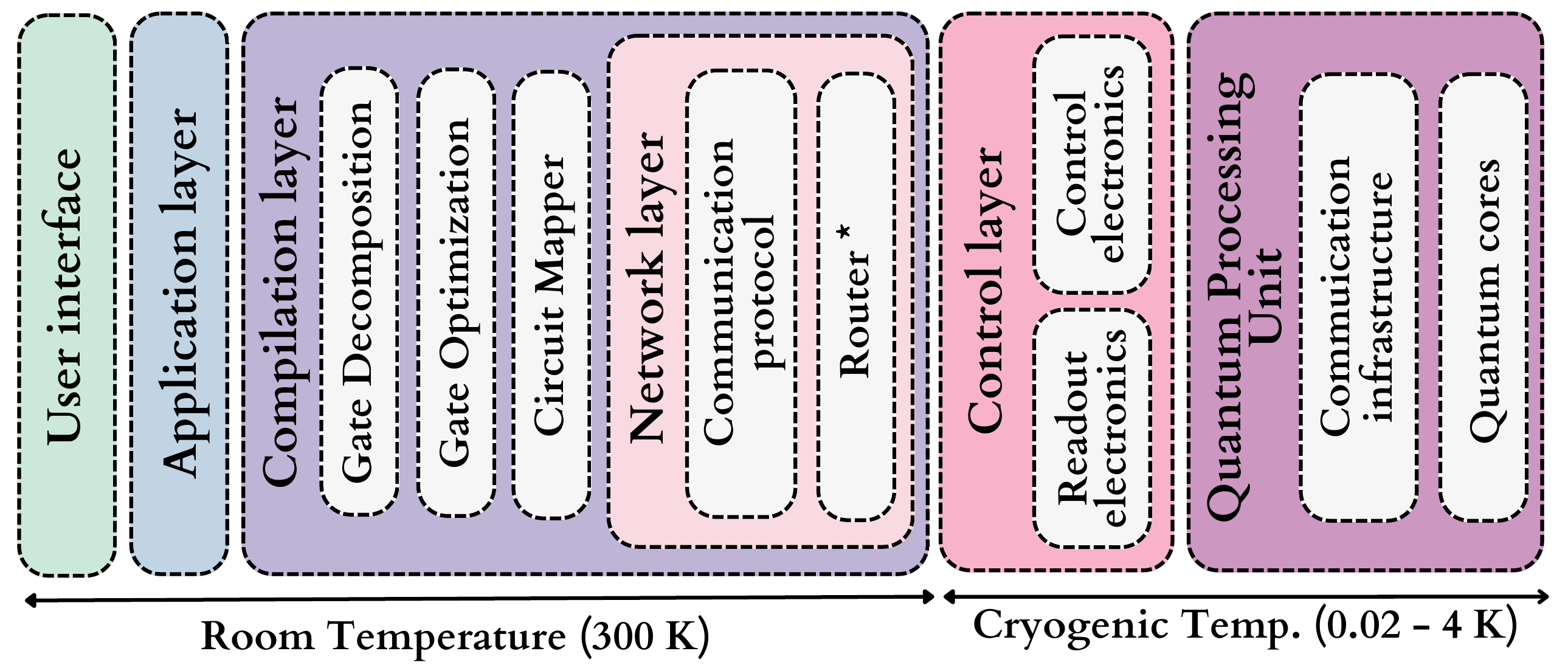}
    \vspace{-0.3cm}
     \caption{Overview of a possible stack for a quantum computer. *In current systems, the routing operations are set at the compilation layer and executed at the network layer, creating a co-dependency between both.} 
     \label{fig:system}
\end{figure}

\section{Background}
\label{sec:background}
In this section, we provide some background on the fundamental properties of quantum computing in Section~\ref{sec:fundamentals}, its implications on the quantum algorithms and the compiling process in Section~\ref{sec:sw}, and a general quantum computing system architecture in Section~\ref{sec:hw}, taking the stack of Figure~\ref{fig:system} as reference.


\subsection{Fundamentals}
\label{sec:fundamentals}

Quantum computing technology leverages particular quantum phenomena to solve computational problems. Essentially, the quantum phenomena underlying quantum computation are the superposition, entanglement, interference, and measurement. We refer the reader to \cite{nielsen00} for a comprehensive explanation of the quantum phenomena applied in quantum computation. The qubit's inherent properties have profound implications in forming a new paradigm in quantum communication as well, fundamentally different from classical communication technologies. For instance, quantum information retransmission is not possible according to the no-cloning theorem \cite{Wootters1982Single} that forbids creating identical copies of an arbitrary unknown quantum state. 

Qubits are typically arranged in arrays forming QPUs and are often placed in cryogenic temperatures to suppress thermal fluctuations in the environment. This is because thermal noise causes qubits to \emph{decohere} over time, this is, to disturb and eventually lose their superposition state and, in a sense, corrupt the entire computation. Hence, cooling qubits to cryogenic temperatures serves to maintain quantum states for a longer period of time or, in other words, to extend their \emph{coherence time}. Additionally, numerous qubit technologies, such as superconducting qubits, are based on materials that become superconducting at cryogenic temperatures. 

Manipulating quantum states for computational purposes requires applying quantum gates, in analogy to classical logic gates, forming the building blocks of quantum algorithms. Single-qubit gates act on one qubit by performing rotations in the computational basis. Two-qubit gates take two qubits as inputs and require both qubits to be in adjacent positions on the processor. Quantum algorithms represent a set of ordered instructions applied on qubits, utilizing phenomena like superposition and entanglement to tackle a given problem. One of the crucial properties of quantum algorithms is quantum parallelism, a remarkable feature that enables the simultaneous evaluation of multiple values of a function. This capability significantly accelerates information processing, allowing quantum computers to explore numerous potential solutions in a relatively brief runtime. 

In the Noisy Intermediate Scale Quantum (NISQ) era of quantum computing \cite{Preskill2018quantumcomputingin}, quantum algorithms are expressed as a sequence of quantum gates applied on qubits, forming quantum circuits. These circuits represent the initial quantum state and the sequence of quantum gates applied to manipulate that state for specific computational tasks. Figure~\ref{fig:second} depicts an example a quantum circuit consisting of four qubits and five gates.


\subsection{Software}
\label{sec:sw}

\begin{figure*}
\vspace{-0.5cm}
\centering
\begin{subfigure}[t]{0.135\textwidth}
    \raisebox{0.45cm}{\includegraphics[width=\textwidth]{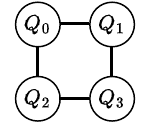}}
    \vspace{-0.8cm}
    \caption{Topology of a quantum computer}
    \label{fig:first}
\end{subfigure}
\hfill
\begin{subfigure}[t]{0.225\textwidth}
     \raisebox{0.15cm}{\includegraphics[width=\textwidth]{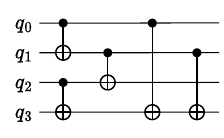}}
    \vspace{-0.8cm}
    \caption{Quantum circuit with 4 virtual qubits and 5 two-qubit gates}
    \label{fig:second}
\end{subfigure}
\hfill
\begin{subfigure}[t]{0.275\textwidth}
    \includegraphics[width=\textwidth]{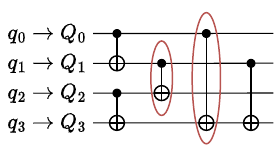}
    \vspace{-0.8cm}
    \caption{Virtual qubits mapped to physical qubits. Unfeasible gates are highlighted}
    \label{fig:third}
\end{subfigure}
\hfill
\begin{subfigure}[t]{0.33\textwidth}
    \raisebox{0.15cm}{\includegraphics[width=\textwidth]{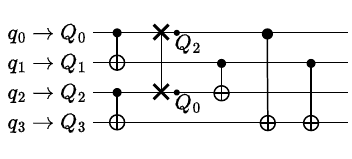}}
    \vspace{-0.8cm}
    \caption{Final mapped circuit. SWAP gates ($\times$) are added to satisfy the topology constaints}
    \label{fig:forth}
\end{subfigure}
\vspace{0.3cm}
\caption{Overview of the process of mapping of a quantum circuit into the topology of a particular quantum computer.}
\label{fig:figures}
\end{figure*}

Acknowledging the availability of powerful software development kits (SDKs) tailored for quantum programming, such as Qiskit \cite{Qiskit} and Cirq \cite{cirqdevelopers_2023_cirq}, is essential. These SDKs provide crucial tools and resources for quantum algorithm development and facilitate seamless interaction with quantum hardware. 

Beyond SDKs, compilers are indispensable components within the quantum computing stack \cite{khammassi2020openql}. As shown in Figure~\ref{fig:system}, they play a pivotal role in translating high-level quantum programming languages into instructions that can be executed on qubits at the processor level. 
Compilers consider the specific properties of the quantum hardware, including qubit control and readout techniques, as well as the processor's topology, i.e. the interconnections of qubits (Figure~\ref{fig:first}). Within the compilation process, gate optimization techniques are applied to enhance the efficiency and performance of quantum circuits, making them less prone to errors. 

One of the most crucial functions of quantum compilers, especially in NISQ computers with limited connectivity between qubits, is quantum mapping \cite{ovide2023mapping}. Its process is illustrated in Figure~\ref{fig:figures}. An initial placement of virtual (or logical) qubits from the circuit to physical qubits of the hardware is performed, storing each quantum state in a qubit. \textbf{Two-qubit gates are applicable only if the involved pair of qubits are adjacent to one another.} Hence, considering the limited physical qubit inter-connectivity, the compiler seeks to add as few routing operations as possible to move quantum states throughout the hardware topology so that each two-qubit gate is executed on adjacent qubits. For future fault-tolerant quantum computers, compilation techniques can include error mitigation strategies such as error correction codes, error-detecting codes, and noise-adaptive circuit optimization to reduce the impact of errors and improve the overall reliability.

\subsection{Hardware}
\label{sec:hw}
The general system architecture of a quantum computer is shown in Figure \ref{fig:comms}. It consists of a host computer containing the circuits to control the execution of a quantum circuit, and a set of QPUs containing the qubit arrays where the quantum circuit is actually executed. Most quantum computers nowadays rely on cryogenic operation, placing the host computer at room temperature and the qubits inside a cryocooler, although recent advances in \textit{hot} qubits \cite{petit2020universal} and cryogenic digital/RF circuits \cite{9772841} are paving the way to having both sub-systems at the same temperature level.

Regardless of its placement, the host computer is in charge of running the compiled circuit, which implies (i) sending the required signals to the quantum computer to control (i.e. apply a quantum gate) or read out (i.e. measure and obtain value) a particular set of qubits at the required instants, and (ii) receiving the result of the readout operations. The nature of these signals depends on the actual qubit technology, as discussed in Section \ref{vertical_comms}, and hence may require the host computer to be equipped with custom digital and analog circuits. This is one of the reasons for using FPGAs.

The QPUs essentially consists of an array of qubits and the necessary circuits to route the control and readout signals to each specific qubit. Depending on the QPU microarchitecture, qubits may be addressed individually \cite{arute2019quantum} or in a clustered way \cite{doi:10.1126/sciadv.aar3960}. The former implies having dedicated cables and pins for each particular qubit, penalizing scalability, while the latter implies the need for extra circuits to route the signals, which affects the quality of the qubits. Another design decision is whether to monolithically integrate all qubits in a single chip, which is complex to achieve as stated beforehand, or to distribute them over multiple chips in a modular setup. In the latter case, extra circuitry is required to connect the chips and orchestrate their operation, as described in Section \ref{horizontal_comms}.

\section{A Primer on Communications in a Quantum Computer}
\label{sec:comms}
\begin{figure}
     \centering
     \includegraphics[width=\columnwidth]{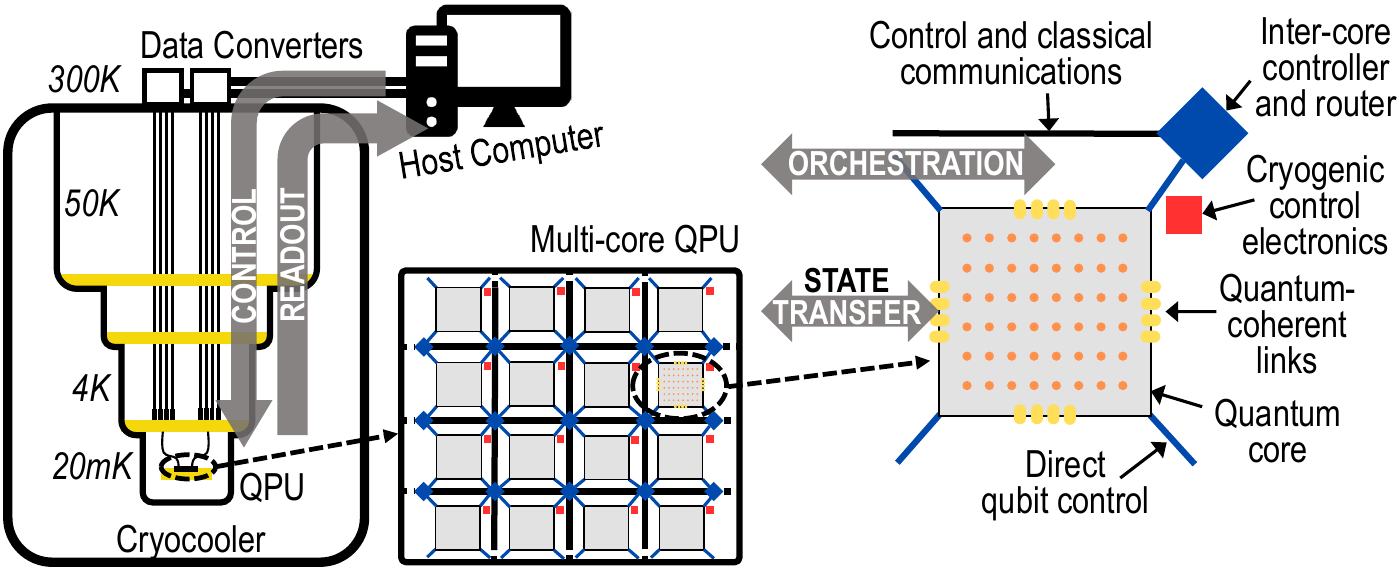}
     \vspace{-0.4cm}
     \caption{Overview of a possible system architecture and communication flows of a modular quantum computer, whose cores may reside in different chips.}
     \vspace{-0.1cm}
     \label{fig:comms}
\end{figure}

Next, we describe the different types of communication occurring within a quantum computer taking as reference the architecture template shown in Figure~\ref{fig:comms}. We distinguish between vertical and horizontal communications, referring to transmissions across temperature levels and within the same temperature level, respectively.

\subsection{\textit{Vertical} Communication} \label{vertical_comms}
According to the general architecture template shown in the Figure~\ref{fig:comms}, vertical communication essentially relates to the \textbf{control} and \textbf{readout} of qubits. Both actions require sending specific signals to the target qubits, whose nature depend on the qubit technology.

Several qubit technologies utilize radio-frequency (RF) signals for the state control and readout. For instance, superconducting qubits are manipulated via precisely shaped RF pulses at the qubit microwave frequency, which is typically 4-8 GHz \cite{Krantz_2019}. 
Similarly, quantum states of trapped-ion qubits are controlled with microwave magnetic fields and RF magnetic field gradients \cite{srinivas2021high} and the nuclear spin of the nitrogen-vacancy centers in diamond can be manipulated using coherent RF pulses \cite{pla2013high} as well. The state readout in quantum dots is commonly performed using RF reflectrometry \cite{schupp2020sensitive, liu2021radio}. As a result, many quantum computers deploy a large set of coaxial cables from the host computer to the QPUs for readout and control.

Typically, a couple of input lines are required for controlling a single qubit, and multiple I/O lines are used for state readout of 1 to 10 qubits \cite{malinowski2023wire}. This highly dense wiring is considered a bottleneck for the scalability of quantum computers, which has prompted the proposal of cryogenic RF switches \cite{potovcnik2021millikelvin} and crossbar architectures \cite{doi:10.1126/sciadv.aar3960} to address more qubits with less cables, turning the coaxial cable array into a more sophisticated interconnect fabric. 

Besides the wiring management problem, there is a bandwidth and power consumption issue as well. With current technologies, readout of qubits demands around 1 Mb/s of bandwidth from the host computer to the QPUs \cite{ganguly2022interconnects}; a simple projection would then foresee a demand exceeding 1 Tb/s in million-qubit systems in the future. Moreover, this flow of data should not exceed the heat dissipation capacity of the cryocooler, which implies the need of transmitting data at a few fJ/bit of energy \cite{ganguly2022interconnects}. As a result of these issues, recent works have also explored the use of alternative technologies such as optical interconnects \cite{lecocq2021control}, or wireless readout schemes \cite{Wang2023} to boost the bandwidth and reduce the power consumption and, hence, thermal disturbance, of the vertical communications. 




\subsection{\textit{Horizontal} Communication} \label{horizontal_comms}
On the horizontal plane, communication is mostly required in the case of implementing a \textit{multi-core} approach, i.e. having the qubits distributed over multiple arrays and/or chips (Figure~\ref{fig:comms}). In that case, the two functionalities requiring communication are the \textbf{qubit state transfer} and the \textbf{qubit transfer orchestration}.

On the one hand, quantum state transfers are essential when two qubits placed in different chips must interact via a two-qubit gate. In that case, either (i) the state of one of the qubits is moved to the other qubit's chip to execute the gate locally, or (ii) a remote gate is applied. Both cases require at least one quantum state transfer, which implies the need of technologies capable of implementing a network of quantum-coherent links between the chips. Currently, multiple communication technologies are under investigation for quantum state transfer at the chip scale, namely: 
\begin{itemize}
\item \textit{Ion shuttling} \cite{kaushal2019shuttlingbased}: is specifically employed in ion-trapped platforms where electromagnetic fields are used to physically transport ions across a chip space deliberately left vacant of qubits, to place them in physical proximity and enable their interaction.
\item \textit{Quantum teleportation} \cite{llewellyn2020chip}: is based on the transfer of state mediated by strongly correlated particle pairs (normally photons) that are sent to the transmitter and receiver. Its use for inter-core communications is still in the early stages of development \cite{10.1145/3477206.3477461}. 
\item \textit{Chip-to-chip interconnects} \cite{c877b567b53e44cda1a0346b58a90d0f}: based on the coupling between two qubits through an appropriate physical medium, which may depend on the physical infrastructure and qubit technology. Alternatives such as low-loss coaxial cables \cite{leung2018deterministic}, superconducting transmission lines \cite{7999762}, and capacitive coupling via resonators \cite{gold2021entanglement} are being investigated. 
\item \textit{Photonic networks} \cite{marinelli2023dynamically}: provide a high degree of connectivity using photons. These photons can be transmitted through waveguides or optical fibers over long distances with low loss and interference, even at room temperature, making them an ideal candidate for inter-core quantum communication. Such an alternative can benefit from the extensive work on photonic NoCs \cite{pasricha2020survey} to create integrated networks for qubit state transfer \cite{zhang2018quantum}.
\end{itemize}

\noindent
On the other hand, we call qubit transfer orchestration to the control operations required to avoid quantum state transfers to conflict with each other. As the number of QPU cores increases, there will be an increasing need for methods to manage functions such as flow control, routing, or synchronization across QPUs, possibly at runtime. This implies the need for fast and efficient transfer of classical data across the chips of the quantum computer, a problem that could leverage the expertise of the NoC community.

\section{A Context Analysis for Interconnect Designers}
\label{sec:context}


In the quest for developing effective interconnect fabrics for modular quantum architectures, it is valuable to perform a context analysis that considers critical aspects related to architectural requirements, physical constraints, and workload characteristics. This section provides an overview of these factors and discusses their implications for interconnect design in quantum computers.

\subsection{Architectural Requirements}
\noindent \textbf{Latency Sensitivity.} Quantum computing's remarkable potential is intricately tied to its ability to execute complex algorithms efficiently. To that end, quantum computers need to be able to execute many quantum gates before qubits \emph{decohere} and lose their state. As a result, all operations (including those involving communication) are extremely latency sensitive, especially in NISQ devices where the coherence time of qubits is very limited. 

\vspace{0.1cm}
\noindent \textbf{Scalability.} To unleash the potential of quantum computers, scaling to a million qubits is required. This will probably require combining scale up and scale out strategies, and making sure that all the layers of the stack are scalable. On the one hand, increasing the number of qubits within a single chip entails balancing the qubit connectivity to maximize the computational power capacity while minimizing undesired interference or crosstalk between neighbouring qubits. Moreover, scalable control electronics 
are equally essential. On the other hand, when interconnecting multiple quantum chips to build more extensive devices, 
the interconnect infrastructure becomes crucial to accommodate a growing number of chips effectively. Efficient routing and scheduling algorithms are necessary to prevent bottlenecks as quantum states move across chips. Lastly, precise synchronization mechanisms are needed for effective inter-chip communications.


\subsection{Physical Constraints}
\noindent \textbf{Cryogenic Temperatures.} As depicted in Section \ref{sec:fundamentals}, most quantum processors operate at cryogenic temperatures. This is a significant physical constraint, as it is hard to ensure millikelvin temperatures across a large space. This may limit the scale of multi-core architectures unless the interconnect infrastructure can remain at higher temperatures and become a quantum-coherent bridge across temperature levels. Also, interestingly, this can be beneficial for the integration of digital/RF circuits in quantum computers due to the improved performance of those circuits under a cryogenic regime.

\vspace{0.1cm}
\noindent \textbf{Ultra-low Power Budget.} The interconnects must not only function effectively under extremely low temperatures, but also align with the rigid power budgets inherent to cryogenic environments. The cryocoolers responsible for maintaining these temperatures introduce specific power consumption and dissipation limitations. Thus, interconnects must be meticulously designed to operate within these confines, ensuring compatibility and functionality.

\vspace{0.1cm}
\noindent \textbf{Variability.} Fabrication process variations impact on any electronic circuit and quantum computers are notably sensitive to process variations \cite{smith2022scaling}. Qubits of the same chip may behave differently and, more relevantly, different chips may be subject to great mismatches. This leads to a diverse range of error profiles that interconnect designers must consider when designing communication strategies. 


\begin{figure*}
\vspace{-0.5cm}
\centering
\begin{subfigure}[t]{0.41\textwidth}
    \raisebox{0.20cm}{\includegraphics[width=0.49\textwidth]{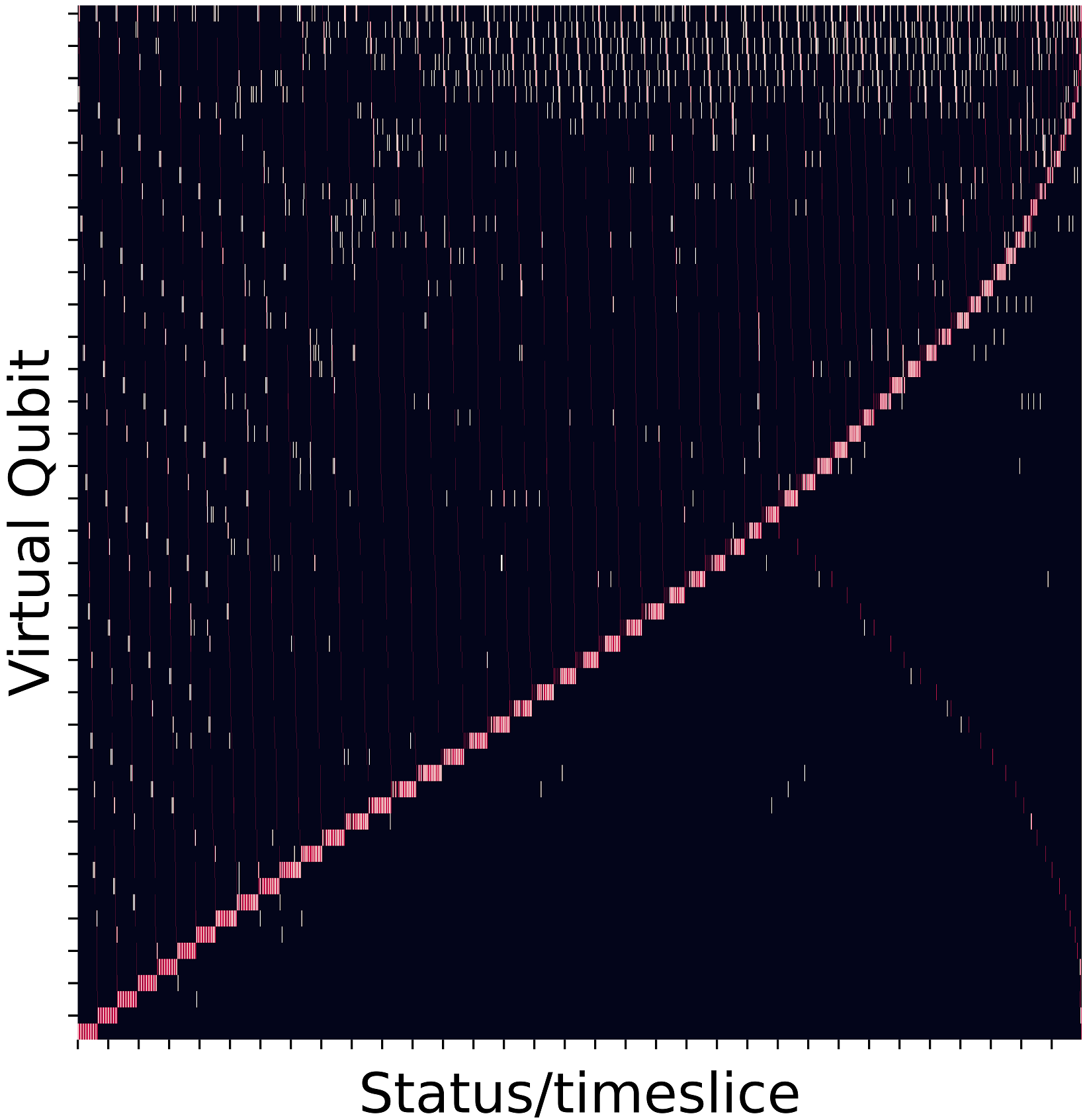}
    \includegraphics[width=0.49\textwidth]{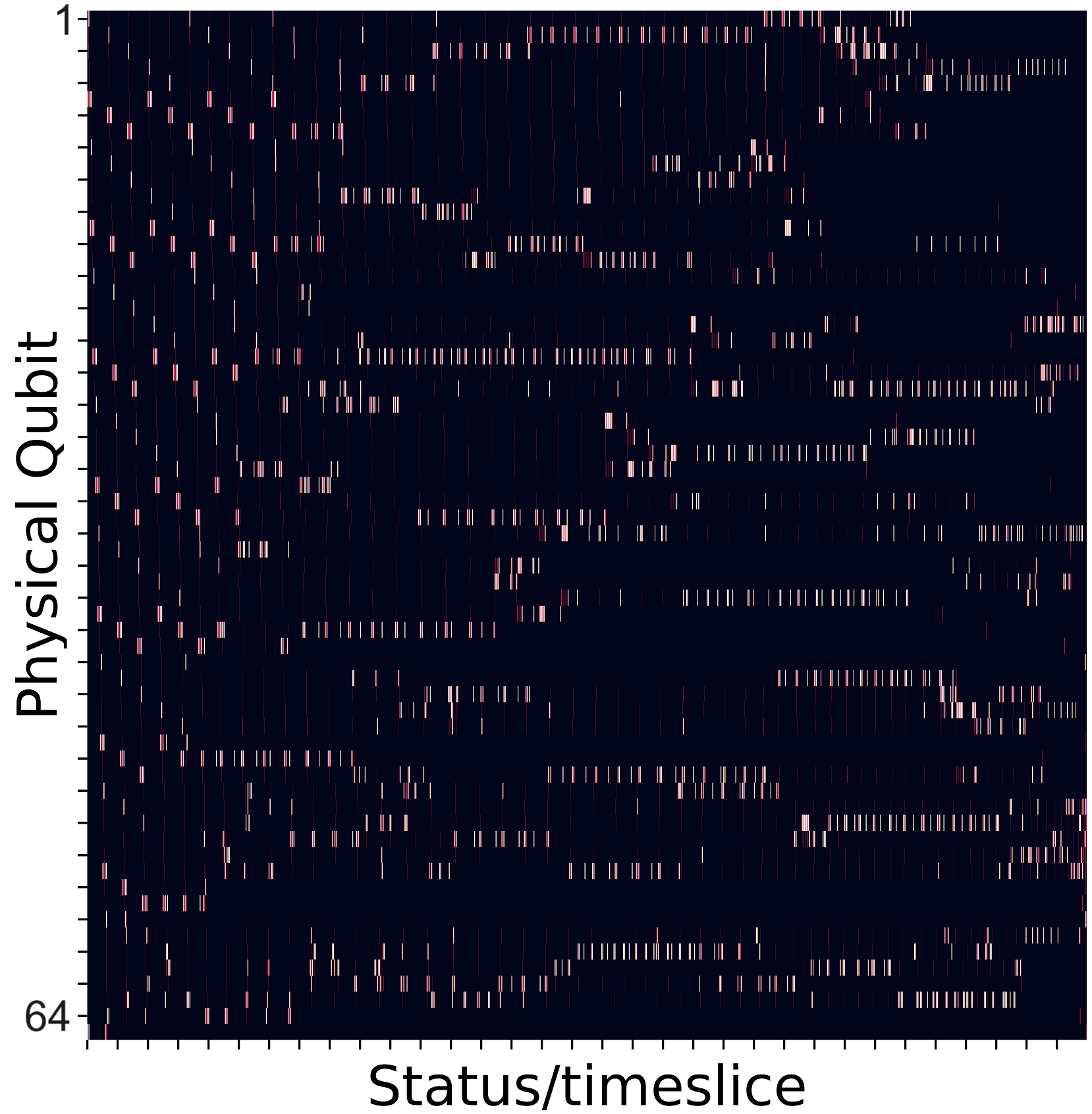}}
    \vspace{-0.35cm}
    \caption{Virtual and physical mapping. Computation, communication, and idling are represented in red, white, and black.}
    \label{fig:virt} \label{fig:phys}
\end{subfigure}
\hfill
\begin{subfigure}[t]{0.24\textwidth}
    \raisebox{0.20cm}{\includegraphics[width=\textwidth]{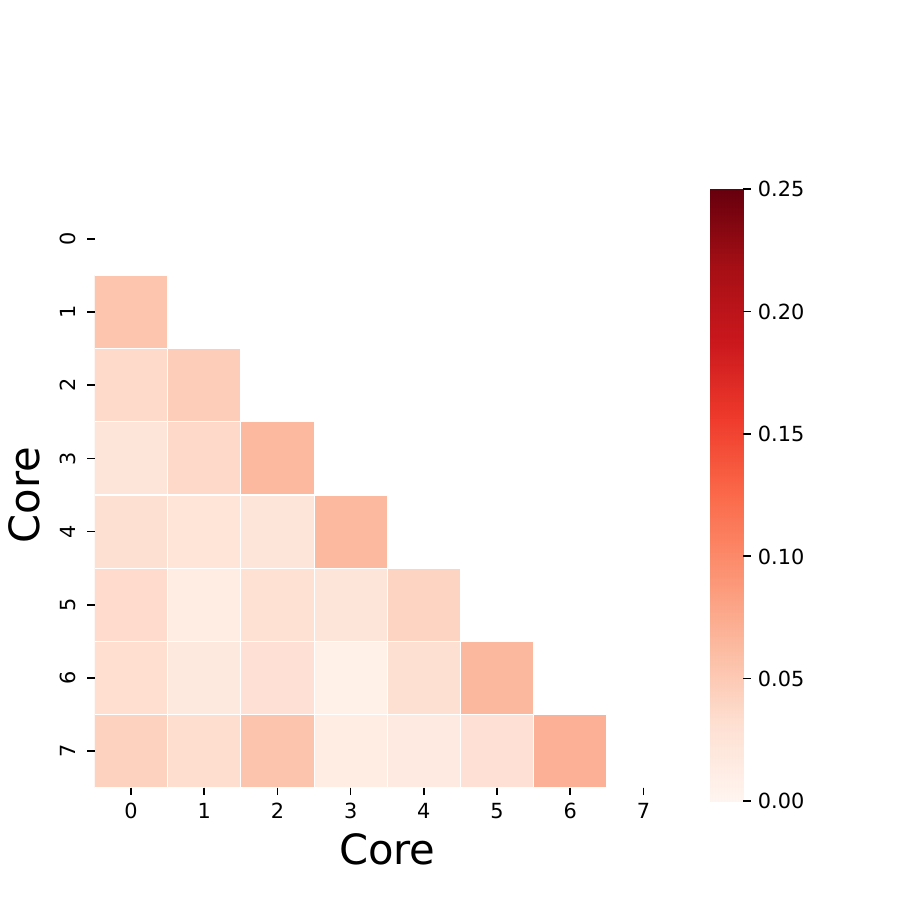}}
    \vspace{-0.75cm}
    \caption{Distribution of inter-core communication operations per core.}
    \label{fig:heatmap}
\end{subfigure}
\hfill
\begin{subfigure}[t]{0.31\textwidth}
    \includegraphics[width=\textwidth]{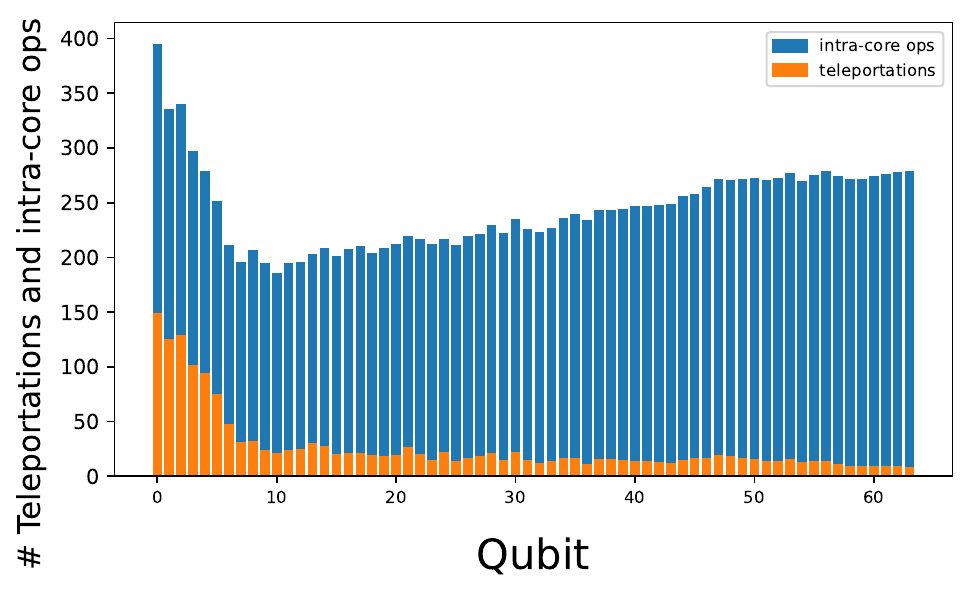}
    \vspace{-0.7cm}
    \caption{Number of teleportation and intra-core operations applied per qubit.}
    \label{fig:telep_ops}
\end{subfigure}
\vspace{0.3cm}
\caption{Qubit traffic analysis of a 64-qubit QFT circuit executed on a 8$\times$8 multi-core architecture with all-to-all connectivity.} 
\label{fig::traffic_heatmap}
\end{figure*}

\subsection{Workload Characteristics}

Characterizing the intra and inter-core workload in modular quantum processors allows to estimate the communications overhead, helping to guide the design of interconnect fabrics for quantum computers. In particular, one can infer the spatio-temporal characteristics of the \emph{vertical} communications by analyzing how quantum gates are spread over the execution of a quantum algorithm, since each gate triggers control signals from the host computer; whereas inspecting which operations occur between qubits placed in different chips offers insight about the \emph{horizontal} communications.

To illustrate the workload characteristics of a multi-core quantum computer, here we follow the methodology of \cite{rodrigo2022characterizing}, which is based on OpenQL \cite{khammassi2020openql}. We compile the 64-qubit Quantum Fourier Transform (QFT) algorithm \cite{nielsen00} for an architecture of 8 cores with 8 qubits each, and then extract various characteristics of the inter-core qubit traffic as depicted in Figure~\ref{fig::traffic_heatmap}. We next provide some insights based on the results of the QFT algorithm, even though the methodology can be used to profile the workload of any algorithm and analyze the differences among them \cite{rodrigo2022characterizing}.


\vspace{0.1cm}
\noindent \textbf{Compilation matters.} By analyzing the circuit mapping traces for the initial virtual circuit and physical circuit in Figure \ref{fig:phys}, we observe the distribution of communication qubits and computation qubits during the program execution. The algorithm structure determines the number of gates, the two-qubit interactions and the inter-core communication requirements. Then, the circuit mapping process seeks to orchestrate those interactions minimizing their overhead, which has an impact on their position (in space) and order (in time).

\vspace{0.1cm}
\noindent \textbf{Spatial distribution.} In Figure \ref{fig:heatmap}, we showcase the inter-core traffic by pairs of cores. This analysis aims to evaluate the uniformity of operation distribution during the program execution time, essentially orchestrated by the compiler. Ideally, favouring a uniform distribution of inter-core communications across the nodes would balance the workload, reduce latency, and improve the efficiency and scalability of a modular processor. This can be achieved by ameliorating the compilation and networking techniques such as initial placement, scheduling, and routing. 

\vspace{0.1cm}
\noindent \textbf{Not all qubits are treated equal.} Another important aspect is the number of communication operations applied to each qubit. This is because each operation decoheres the involved qubits, so that uneven distributions may reduce the validity of the entire algorithm. As a result, we also evaluate the distribution of operations by determining the number of teleportation and intra-core computation operations applied per qubit. As displayed in Figure~\ref{fig:telep_ops}, the distribution of both types of operations depends on the algorithm structure that dictates the gate sequence applied on each qubit, as well as the number of communication operations necessary to comply with the execution of the program in a multi-core setting. This calls for methods to establish priority (in a sort of QoS fashion) for qubits subject to a higher number of operations, so that the overall load is homogeneously distributed across all qubits. 


\section{Discussion}
\label{sec:challenges}

As outlined in previous sections, building robust modular quantum processors is highly contingent on designing and implementing an efficient interconnect fabric for communication across the quantum chips and to the host computer. To this aim, several challenges need to be tackled across the full stack. Here, we discuss several overarching challenges and possible ways to address them. 

\vspace{0.1cm}
\noindent \textbf{Compile-time vs Run-time.} 
The very stringent requirement of latency (rooted in the limited computation times given the qubit decoherence) renders the scaling of quantum computers extremely difficult. The former calls for executions completely driven by compile-time mapping and routing, which is the technique currently used in quantum computing. However, existing compilers use algorithms that hardly scale to millions of qubits, suggesting that some functionalities (including qubit routing) might need to be executed in run-time. Therefore, developing techniques to minimize the communication latency is instrumental for the development of modular quantum processors. One might find inspiration in NoCs for real-time systems \cite{hesham2016survey}, fixing deadlines and assigning priorities based on the time left for the qubits to decohere. In any case, qubit technology has to be taken into account as it dictates the qubit decoherence times, topology, and others.


\vspace{0.1cm}
\noindent \textbf{Co-Design Methodologies.} Abstraction layers in communications and computing appear when there are enough resources to support it. Alas, currently quantum computing cannot afford the loss of efficiency typically associated to abstraction layers, given its extremely stringent latency and power budgets. Instead, quantum computing requires a coordinated effort between hardware and software, also because the performance of quantum algorithms often depends on the specific characteristics of the hardware. Then, it is expected that co-design will be exploited in the multi-core quantum computers. Clear examples are mapping and resource allocation techniques, whose success greatly depend on aspects such as the algorithm structure, qubit connectivity, gate fidelities, or the or the cost of moving qubits within and across chips. Similarly to recent trends in co-design in deep learning \cite{lie2023cerebras}, one could conceive techniques where the algorithms, the compiler, and the runtime techniques are co-designed with the interconnect.



\vspace{0.1cm}
\noindent \textbf{Simulation Tools.}
Multiple software tools have been presented in recent years to simulate quantum computers on the different layers of the stack. Functional simulators such as QX \cite{khammassi2017qx}, Qiskit's Aer \cite{Qiskit}, or mpiQulacs \cite{Imamura2022mpiQulacsAD} simulate the output of a quantum algorithm in ideal conditions or under specific noise models, yet at the expense of huge computational and memory requirements, whereas other simulators have instead focused on other aspects such as the control system surrounding the quantum computer \cite{riesebos2022functional}. However, simulators of modular quantum processors integrating various inter-core communication protocols are still missing, mostly due to the lack of quantum-coherent link models and because the impact of noise (including that of the quantum-coherent links) is not fully understood \cite{resch2021benchmarking}. To bridge this gap, Quantum Internet communication simulators such as NetSquid \cite{coopmans2021netsquid}, together with multi-chip NoC simulators \cite{bharadwaj2020kite}, could be adapted to become the core of a unified simulation framework for modular quantum processors.  


\section{Conclusion}
\label{sec:conclusion}
Multi-core quantum architectures are a promising approach to enable scalable quantum computing. However, the realization of such approach towards the million qubit barrier poses a myriad of challenges, many of which relate to communications from/to/within a quantum computer. As the present context analysis has shown, such problems are reminiscent to those found in classical processors and, therefore, conventional knowledge from the NoC community may be inspiring; yet the unique blend of ultra-low latency requirements, cryogenic operation, and sensitivity to process variations renders the development of interconnect fabrics for quantum computers particularly challenging. We conclude pointing at the development of novel run-time techniques and simulation tools with a co-design approach as key aspects to advance in this formidable quest.




\appendix


\end{document}